\documentstyle[psfig,conf_iap]{article}
\begin{document}
\heading{%
%Begin Heading
%
Measuring galaxy biasing with gravitational weak lensing
%
%End Heading
} 
\par\medskip\noindent
\def \vc #1{{\textfont1=\bolditalics \hbox{$\bf#1$}}}
\def\kg{{\rm \bf k}}
\author{%
%Begin Author names
Ludovic Van Waerbeke
%End Author names
}
\address{%
Canadian Institut for Theoretical Astrophysics, 60 St George Str.,
M5S 3H8 Toronto, Ontario, Canada
}

\begin{abstract}
Gravitational weak lensing by large scale structures is view as a tool
to probe the bias relation between the mass and the light distributions. It is
explained how a particular statistic can be used to deproject the 2D mass
distribution observed by weak lensing, in order to probe the bias at
a given redshift and a given scale. Dependence with the cosmology is written,
and observational issues are pointed out, like the importance of the redshift
distribution of the galaxies. A signal to noise analysis shows that the scale
dependence of the bias can be measured with a rather modest lensing survey
size.
\end{abstract}
\section{Weak lensing and galaxy catalogues}
Gravitational weak lensing by large scale structures is known to be a promising
aproach to probe the mass distribution in the Universe. Gravitational lensing
by cluster of galaxies already gave new insights about the cluster mass and
their 2D mass distribution whereas observation of lensing by large scale
structure is only at its beginning (\cite{MF98}). Its potential scientific impact on
the knowledge of the cosmological parameters and the mass power spectrum
pionnered by \cite{BSBV}, \cite{G67}, \cite{K92}, \cite{M91},
has been developped by many authors (see \cite{BvWM}, \cite{JS97},
\cite{K98}, \cite{V96}). One may also
be interested in comparing the mass maps reconstructed from weak lensing and the
galaxy catalogues (i.e. the mass and the light distributions) in
order to probe the so-called bias relation.
Galaxy catalogues (via high order statistic) and velocity fields being the
standard ways to probe the bias, a method able to {\it see} directly the mass,
like
gravitational lensing, may play a role. In the following some clues about the
measurement of bias using gravitational lensing will be exposed, most of
them are detailed in \cite{VW98}.

Biasing theory has known recently a raise of interest because of the growing
evidences that it is not trivial. However, in order to give some
insights on the method, local linear biasing relation is used as a basis along
this paper, and the extension to non trivial biasing is let
for a forthcoming paper \cite{VW98_b}.

The basic idea is to probe the bias by measuring the correlation ${\cal C}$
between the projected density contrast of a set of foreground
galaxies and the shear of background galaxies.
The difficulty with the gravitational lensing is that it gives the projected
mass along the line of sight, whereas the 3D mass field is required in order
to constrain the bias, in particular if we wish to investigate its redshift
evolution. Fortunately the $M_{\rm ap}$ statistic introduced by
\cite{SvWJK} gives the possibility to deproject the observed mass distribution.
To understand how it is possible, let us write the correlation function
${\cal C}$ explicitly \cite{S98},
\begin{equation}
{\cal C}=\langle M_{\rm ap}(\theta_{\rm c}){\cal N}(\theta_{\rm c})\rangle
=3\pi \Omega b \int {\rm d}w
{p_{\rm f}(w) g(w)\over a(w) f_K(w)}\int {\rm d}s~s P\left({s\over f_K(w)},
w\right) I^2(s\theta_{\rm c})
\label{MN_def}
\end{equation}
$M_{\rm ap}(\theta_{\rm c})$ is the convergence smoothed with a compensated
filter of size $\theta_{\rm c}$, and ${\cal N}(\theta_{\rm c})$ is the
projected density contrast of the foreground galaxies smoothed with the same
filter. $b$ is the linear bias parameter, $w(z)$ is the comoving distance to
a redshift $z$, $f_K$ is the
comoving angular diameter distance, and $P$ is the time-evolving 3-D mass power
spectrum. $\Omega$ is the density parameter, and $a$ is the cosmic expansion
factor.
The function $g(w)=\int_w^{w(\infty)}~{\rm d}w'~p_{\rm b}(w')~f_K(w'-w)/f_K(w')$
depends on the redshift distribution of the sources $p_{\rm b}(w)$ and
$p_{\rm f}(w)$ is the redshift distribution of the foreground galaxies.
$I(s\theta_{\rm c})$ is the fourier transform of the compensated filter,
which is very localised in the fourier space around the angular
scale $\theta_{\rm c}$. If in addition a narrow
redshift distribution is used for the foreground galaxies then the integrals
in Eq. (\ref{MN_def}) are almost reduced to a point in the $(w,s)$ plane,
which is equivalent to a deprojection (a single angular scale $s$ is
observed at a given redshift $w_f$).

\section{Probe of the bias}
\begin{figure}
\centerline{\hbox{
\psfig{figure=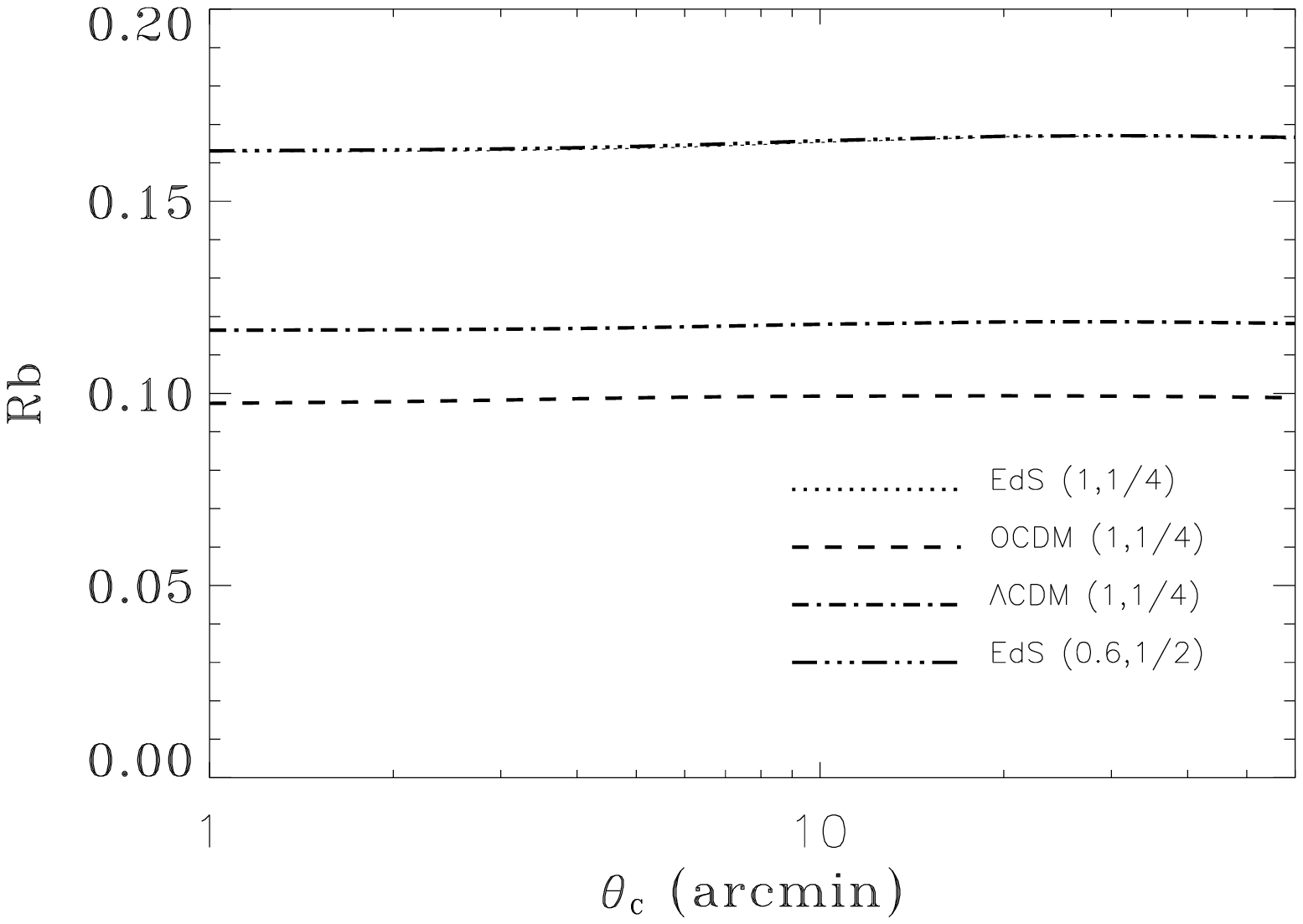,height=5.cm}
\psfig{figure=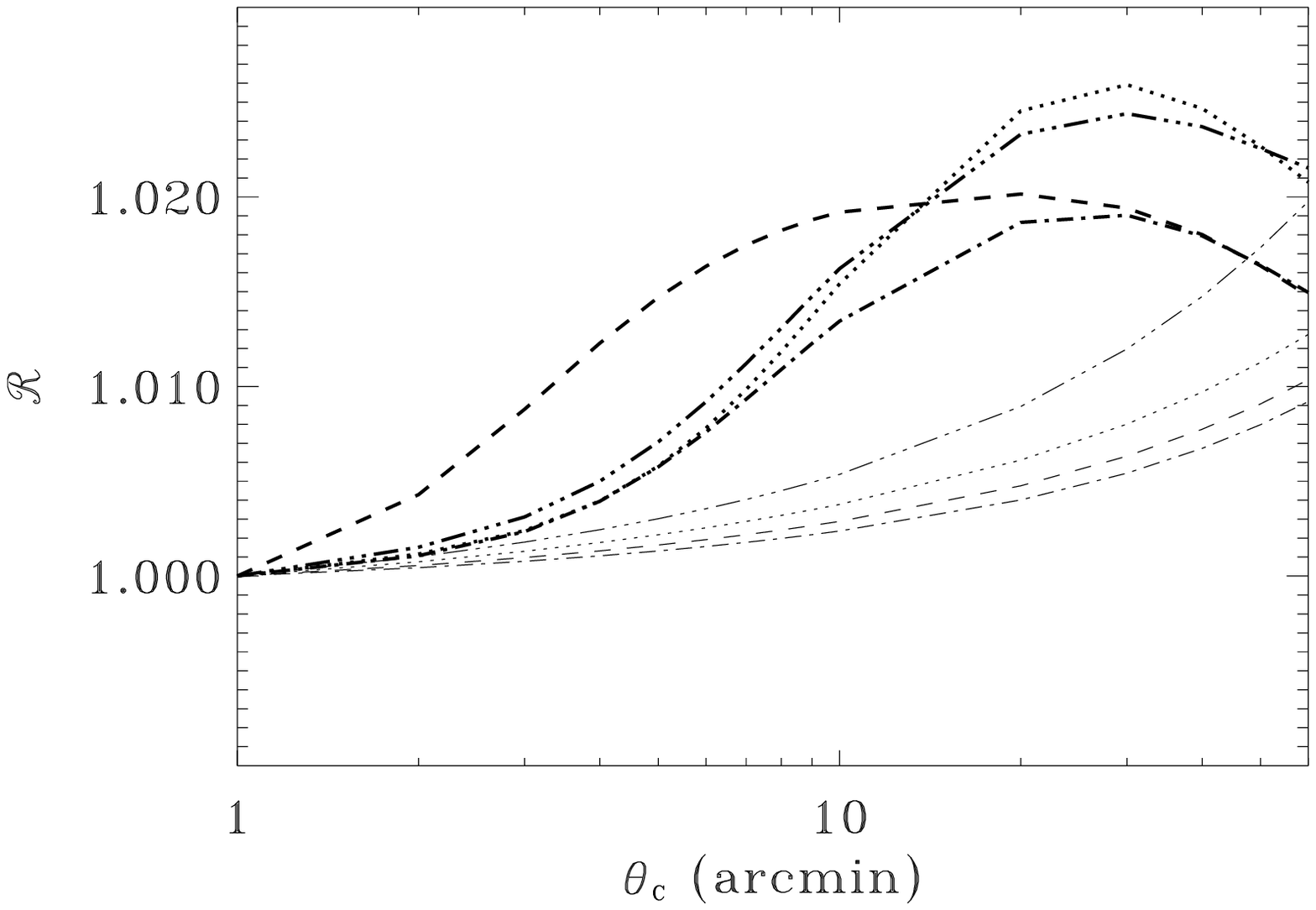,height=5.cm}
}}
\caption{\label{script_R.ps}}{The left plot shows the value of
$R_{\theta_{\rm c}}b$ and the right plot the ratio ${\cal R}=R_{\theta_{\rm c}}/
R_{1'}$. Different linestyle correspond to different cosmological models
as indicated on the left plot, thin lines are for the linear regime and
thick lines for the non-linear regime. CDM-like power spectrum has been used.
}
\end{figure}

A useful estimator to probe the bias, independent on the power spectrum
normalisation is,
\begin{equation}
R_{\theta_{\rm c}}={\langle M_{\rm ap}(\theta_{\rm c}){\cal N}(\theta_{\rm c})\rangle\over
\langle {\cal N}^2(\theta_{\rm c})\rangle},
\label{R_def}
\end{equation}
where $\langle {\cal N}^2(\theta_{\rm c})\rangle $ is the dispersion of the smoothed projected
density contrast of the foreground galaxies,
\begin{equation}
\langle {\cal N}^2(\theta_{\rm c})\rangle =2\pi b^2 \int {\rm d}w {p_{\rm f}^2(w
)\over f_K^2(w)} \int {\rm d}s~s~P\left({s\over
f_K(w)},w\right) I^2(s\theta_{\rm c}).
\label{N2_def}
\end{equation}
In the case of linear theory, for a power law power spectrum and for a narrow
foreground redshift distribution (located around $w_f$) it is easy to show
from Eq.(\ref{MN_def}) and
Eq.(\ref{N2_def}) that Eq.(\ref{R_def}) can be approximated by,
\begin{equation}
R_{\theta_{\rm c}}\simeq{3\over 2} {\Omega\over b} {g(w_{\rm f})
f_K(w_{\rm f}) p_{\rm f}(w_{\rm f})\over a(w_{\rm f}) \int dw~p_{\rm f}^2(w)},
\label{R_approx}
\end{equation}
which is a number depending only on the cosmological parameters, the biasing
factor $b$ and the foreground and background redshift distributions. It does not
depend on the power spectrum index and the smoothing angle $\theta_{\rm c}$.
It turns out that this is still true for a general power spectrum even in the
non linear regime. The reason for this is that if the redshift integration
in Eq.(\ref{MN_def}) and Eq.(\ref{N2_def}) are narrow enough, a general power
spectrum can be approximated {\it locally} as a power law because of the narrow
function $I(s\theta_{\rm c})$. {\it Locally} here means in the $(w,s)$ space.
An illustration of this is given if figure \ref{script_R.ps} where
$R_{\theta_{\rm c}}$ and ${\cal R}=R_{\theta_{\rm c}}/R_{1'}$ are
plotted versus the smoothing angle $\theta_{\rm c}$ for different cosmological
models, and with the narrow foreground \footnote{The background redshift
distribution is $p_b(z)\propto z^2~\exp (-z^{1.5})$} redshift distribution
$p_f(z)\propto z^5~\exp -(z/0.4)^6$ well localised around $z_f=0.4$.
The power spectrum evolution has been extended into the
non-linear regime via the Peacock \& Dodds formula in \cite{PD96}. Figure
\ref{script_R.ps} shows that $R_{\theta_{\rm c}}$ is has a constant value
to a percent accuracy over the scale range $[1',60']$ which only depends
on the cosmological parameters, the bias (at the redshift of the foregrounds)
and the redshift distributions whatever
the power spectrum. Calculations with various power spectra shows 
that $R_{\theta_{\rm c}}$ changes by less than $3\%$.
The $R_{\theta_{\rm c}}$ values fitted to the $\Omega$ dependence (assuming
a zero cosmological constant) gives,
\begin{equation}
R_{\theta_{\rm c}}\simeq 0.162{\Omega^{0.42}\over b},
\label{R_fit}
\end{equation}
where the proportionality constant and the exponent both depend on the redshift
distributions according to Eq.(\ref{R_approx}). Conjointly used with the
velocity fields or any independent determination of $\Omega$, Eq.(\ref{R_fit})
allows to determine the bias at the redshift of the foreground galaxies (here
roughly $0.4$) and at the scale $\theta_{\rm c}$ from weak lensing data provided
that the
redshift distributions are known. Preliminary investigations show that only
the mean and the variance of these distributions are important even if they 
are skewed.

The quantity ${\cal R}$ defined above gives a direct estimate of the scale
dependence of the bias, independent on the cosmology, the power spectrum
and the redshift distributions. Strictly speaking it gives the ratio of the
fourier components of the bias at two different scales provided that the bias
can be expressed as a convolution, $\tilde \delta_g(\kg)=
\tilde b(\kg)\tilde\delta(\kg)$ (linear non-local non-stochastic biasing sheme).
%
%\begin{figure}
%\centerline{\vbox{
%\psfig{figure=errors.ps,height=4.cm}
%}}
%\caption{\label{errors.ps}}{Expected $1\sigma$ error bars on the measurement of
%${\cal R}=R_{\theta_{\rm c}}/ R_{1'}$ from a lensing survey of $5\times 5$
%degrees.
%lensing inversion.
%}
%\end{figure}
%
A signal to noise analysis shows that
assuming that $\langle {\cal N}^2(\theta_{\rm c})\rangle$ is perfectly known
(from the future big galaxy surveys for instance),
a lensing survey of $5\times 5$ degrees
(with 4 hours exposure depth) should be able to detect a bias scale
variation of $20\%$ or more in the $[1',10']$ scale range.

\section{Conclusion}

Gravitational lensing can probe the bias relation between mass and light
distributions. The simultaneous use of a narrow foreground redshift distribution
and a compensated filter permits to {\it deproject} the observed mass
distribution and to recover the redshift and scale dependence of the bias
using Eq.(\ref{R_fit}) and the estimator ${\cal R}$. Redshift distorsion
effects are not a problem in this method since the foreground redshift range
is large enough do dilute these effects, and on the other hand, it is small
enough to assume that a given angular scale corresponds to the physical scale
at which the bias is probed. Of course, the statistical bias properties are
assumed to remain unchanged within the foreground galaxies redshift interval.

Obviously the biasing sheme adopted in this paper
is too simple and should be extended to non-linear, stochastic and
non-local biasing. One can hope to use weak lensing jointly with galaxy
catalogues and velocity fields to separate these
features and try to build a consistent bias picture.

\acknowledgements{Many thanks to F. Bernardeau, Y. Mellier, and P. Schneider
for regular and enthousiastic discussions. I thank the {\it Programme National
de Cosmologie} for financial support.}

\vfill
\end{document}